\documentclass[11pt,a4paper]{article}
\usepackage[utf8]{inputenc}
\usepackage[english]{babel}
\usepackage{amsmath}
\usepackage{amsfonts}
\usepackage{amssymb}
\usepackage{graphicx}
\usepackage{listings}
\usepackage{float}
\usepackage{booktabs}
\usepackage{geometry}
\usepackage{subfig}
\usepackage{ulem}
\usepackage{spalign}
\usepackage[none]{hyphenat}

\usepackage{vmargin}

\setpapersize{A4}
\setmargins{3cm}       
{1.5cm}                        
{15.25cm}                      
{23.42cm}                    
{10pt}                           
{1cm}                           
{0pt}                             
{2cm}                           

\usepackage{graphicx}
\title{Linearization of a dual-parallel Mach-Zehnder modulator using optical carrier band processing}
\author{Luis Torrijos-Morán, Cristina Catalá-Lahoz, Daniel Pérez-López, Li Xu, Wuang Tianxiang and Diego Pérez-Galacho\\
\vspace{0.75cm}
\small{Photonics Research Labs, Universitat Politècnica de València, 46022 Valencia, Spain. }\\
\small{iPronics Programmable Photonics, 46022 Valencia, Spain.}\\
\small{Central Research Institute, Huawei Co. LTD.}\\
\small{luistm@upv.es}} 

\makeatletter         
\def\@maketitle{
\begin{center}
{\Huge \bfseries \sffamily \@title }\\[4ex] 
{\Large  \@author}\\[4ex] 
\@date\\[8ex]
\end{center}}
\makeatother

\begin{document}
\normalem
\maketitle
\sloppy


\vspace{-0.25cm}

The linearization of a microwave photonic link based on a dual-parallel Mach-Zehnder modulator is theoretically described and experimentally demonstrated. Up to four different radio frequency tones are considered in the study, which allow us to provide a complete mathematical description of all third-order distortion terms that arise at the photodetector. Simulations show that a complete linearization is obtained by properly tuning the DC bias voltages and processing the optical carrier band. As a result, a suppression of 17 dB is experimentally obtained for the third-order distortion terms, as well as a SDFR improvement of 3 dB. The proposed linearization method enables the simultaneous modulation of four different signals without the need of additional radio frequency components, which is desirable to its implementation in integrated optics and makes it suitable for several applications in microwave photonics.

\section{Introduction}

Microwave photonics links (MPL) provide several advantages in terms of bandwidth, low loss, lightweight and immunity to electromagnetic interference compare to conventional electrical links, which makes them suitable for radio frequency (RF) applications such as wireless communications, radio-over-fiber and antenna remoting, among others \cite{Capmany2007,Yao2009}. However, nonlinear distortion arises as the main limitation factor of MPLs, wherein the electro-optic modulator is the one that contributes the most to produce this effect. Nonlinear distortion can be classified as harmonic distortion (HD) or intermodulation distortion (IMD) and the spurious free dynamic range (SFDR) is the figure of merit normally used to measured IMD. \cite{Cox2006}. HD lie away from the fundamental signals so that they can be filtered out. In contrast, the third-order intermodulation distortion (IMD\textsubscript{3}) is so close to the fundamental frequencies and it cannot be easily removed using filters \cite{Ferreira2009}.

To improve the linearity of the MPL, first approaches were proposed decades ago by using additional modulation structures to produce complementary IMD\textsubscript{3} terms so that they cancel the primary modulator distortion at the photodetector (PD) \cite{Ackerman1999}. These schemes are designed using complementary Mach-Zehnder modulators (MZM) either in series or parallel to provide broad-band linearization but at the expense of increasing the structural complexity, size and power consumption. Another linearization technique is based on optical spectrum processing in order to identify the main contributors of IMD\textsubscript{3} and cancel them by tuning the carrier and sidebands of the modulated signal. This was first proposed for intensity-modulation where it was demonstrated that only a phase shift in the optical carrier band is needed to suppress IMD\textsubscript{3} terms in a MZM \cite{Zhang2012b,Zhang2012,Cui2013,Zhu2016b}. This method was later extended to phase modulators (PM) providing the advantages of low loss and simplicity but requiring a more complex optical spectrum processing since the phase and amplitude of the sidebands must be also considered \cite{Li2013b,Li2014,Wu2019,Liu2020}.

In parallel, more complex structures such as dual-parallel MZM (DPMZM) have also been investigated for linearization by adjusting the power between the electrical drive signals \cite{Korotky1990,Zhou2016}, input and output optical power splitting ratios \cite{Zhu2009,Kim2011}, optimizing the working points of single-drive configurations \cite{Li2010}, or properly designing electrical phase shifters to have an active control of the RF inputs \cite{Li2013,Jiang2015}. Likewise, similar DPMZM are linearized using complementary IMD\textsubscript{3} terms at each MZM \cite{Gu2019}, or even another DPMZM in parallel as a polarization-multiplexing configuration \cite{Zhu2016}. Nevertheless, in all aforementioned DPMZM the underlying idea stems from introducing a certain predistortion in the RF domain to compensate existing non-linearities in one of the MZM, which reduces the bandwidth limited by the electrical components. So far, an exhaustive analysis of the DPMZM linearization by means of optical processing remains unexplored as well as its implementation without additional RF techniques.

In this paper, we present the linearization of a DPMZM by processing the carrier band in the optical domain. To this end, both MZM are configured in push-pull configuration so that no additional RF components are needed. In addition, a four tone test is considered in the linearization scheme which allows us to demonstrate that two different RF signals can be modulated at the same time in each MZM without third-order distortion. Experimental measurements are also provided, showing a good agreement with theoretical predictions and simulations. The proposed work extends the use of optical processing for the linearization of a more complex structure such as DPMZM with interesting results and applications.

\section{Operation principle}

The proposed MPL is shown in Fig. 1. It is composed by a DPMZM with both individual MZM in push-pull configuration,  —i.e., a phase shift of $\pi$ between driving RF signals applied to each of the MZM arms. In this scheme, the optical carrier is modulated by two different RF signals: $\omega_{1,2}$ and $\omega_{3,4}$ in the case of the upper and lower MZM, respectively, so that a four-tone test is considered. We can define the optical field at the output of the upper and lower MZM, $E_{1}$ and $E_{2}$ respectively, as:
\begin{equation}
\begin{split}
&E_{1}(t)=-j \sqrt{P_i} e^{j \omega_c t} \sqrt{k-k^2}\\
&\cdot \left[e^{j\phi_{1}} e^{jm \sin \omega_1t} e^{jm \sin \omega_2t} + e^{-jm \sin \omega_1t} e^{-jm \sin \omega_2t} \right]
\end{split}
\end{equation}
\begin{equation}
\begin{split}
&E_{2}(t)=-j \sqrt{P_i} e^{j \omega_c t} \sqrt{k-k^2}\\
&\cdot \left[e^{j\phi_{2}} e^{jm \sin \omega_3t} e^{jm \sin \omega_4t} + e^{-jm \sin \omega_3t} e^{-jm \sin \omega_4t} \right]
\end{split}
\end{equation}
where $P_i$ is the input optical power, $\omega_c$ the optical carrier angular frequency, $k$ the coupling coefficient, $\phi$ the phase shifts that control the bias point of each MZM and $m$ the modulation index. Note that the exponent of the second term is considered negative due to the push-pull configuration of the MZM. Therefore, the output field of the complete DPMZM in turn can be calculated as the sum of Eq. (1) and (2) as
\begin{equation}
E_{out}(t)=\sqrt{k-k^2} \left[e^{j\phi_{3}}E_{1}(t)+E_{2}(t)\right].
\end{equation}

For the sake of simplicity and in order to expand the field into Bessel functions of the first kind, we must first define these two variables
\begin{equation}
S_1(m)=\sum_{n=-\infty}^{+\infty} J_n(m)e^{jn\omega_1t}\ \sum_{k=-\infty}^{+\infty} J_k(m)e^{jk\omega_2t}
\end{equation}
\begin{equation}
S_2(m)=\sum_{n=-\infty}^{+\infty} J_n(m)e^{jn\omega_3t}\ \sum_{k=-\infty}^{+\infty} J_k(m)e^{jk\omega_4t}
\end{equation}
so that the modulated optical spectrum of the field in Eq. (3) can be rewritten as
\begin{equation}
E_{out}(t)\propto e^{j\phi_{3}} \left[ e^{j\phi_{1}} S_1(m) + S_1(-m) \right] + \left[ e^{j\phi_{2}} S_2(m) + S_2(-m) \right].
\end{equation}

\begin{figure*}[b]
\centering\includegraphics[scale=1.25]{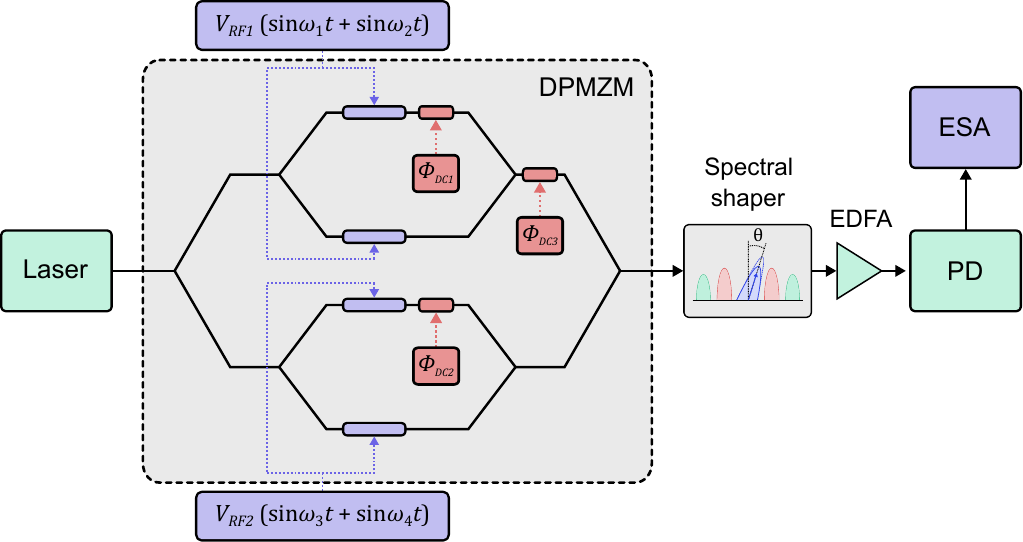}
\caption{Schematic of the presented MPL based on the linearization of a DPMZM in a push-pull configuration under a four-tone test. EDFA means erbium-doped fiber amplifier, PD photodetector and ESA electronic spectrum analyzer.}
\end{figure*}

The modulated optical spectrum at the DPMZM output is therein processed by the spectral shaper to impose a certain amplitude suppression and phase shift on the optical carrier band (OCB). Considering first and second order sidebands ($\pm$OSB and $\pm$2OSB) in the optical spectrum, there are four beating products (BPs) that produce IMD\textsubscript{3} at the photo-diode, as it is detailed in Fig. 2a. The first one is the beating between the frequency component from the OCB: $\omega_c$ with the components from the $\pm$OSB: $\omega_c+2\omega_{1,2}-\omega_{2,1} (\omega_c-2\omega_{1,2}+\omega_{2,1})$ and $\omega_c+2\omega_{3,4}-\omega_{4,3} (\omega_c-2\omega_{3,4}+\omega_{4,3})$. This BP produces direct IMD\textsubscript{3} terms that only depend on one of the RF frequencies, —e.g., $2\omega_{1}-\omega_{2}$. The second BP in turn includes the mixing between the component from the OCB: $\omega_c+\omega_{1,2}-\omega_{2,1}$ and $\omega_c+\omega_{3,4}-\omega_{4,3}$ with the component from the $\pm$OSB: $\omega_c+\omega_{1,2} (\omega_c-\omega_{1,2})$ and $\omega_c+\omega_{3,4} (\omega_c-\omega_{3,4})$. This BP produces direct IMD\textsubscript{3} terms as before, but also crossed IMD\textsubscript{3} terms that depend on both RF\textsubscript{1} ($\omega_{1,2}$) and RF\textsubscript{2} ($\omega_{3,4}$) frequencies —e.g., $\omega_{1}-\omega_{2}+\omega_{3}$. The third BP involves the mixing of the component from the $\pm$OSB: $\omega_c-\omega_{2,1} (\omega_c+\omega_{2,1})$ and $\omega_c-\omega_{4,3} (\omega_c+\omega_{4,3})$ with the component from $\pm$2OSB $\omega_c+2\omega_{1,2} (\omega_c-2\omega_{1,2})$ and $\omega_c+2\omega_{3,4} (\omega_c-2\omega_{3,4})$. This BP produces both direct and crossed IMD\textsubscript{3} terms —e.g., $2\omega_{1}-\omega_{2}$ and $2\omega_{1}-\omega_{3}$, depending on whether the beating is between the same RF signal or not, respectively. Finally, there is a fourth BP between the component from the $\pm$OSB: $\omega_c-\omega_{2,1} (\omega_c+\omega_{2,1})$ and $\omega_c-\omega_{4,3} (\omega_c+\omega_{4,3})$ with the component from $\pm$2OSB: $\omega_c+\omega_{1,2}+\omega_{2,1} (\omega_c-\omega_{1,2}-\omega_{2,1})$ and $\omega_c+\omega_{3,4}+\omega_{4,3} (\omega_c-\omega_{3,4}-\omega_{4,3})$. This BP produces only crossed terms —e.g., $\omega_{1}+\omega_{2}-\omega_{3}$. It is worth noting that the fourth BP also produces coefficients $I_1$ when the mixing is between frequency components of the same RF signal  —e.g., $\omega_{1}+\omega_{2}-\omega_{2}=\omega_{1}$.

Once the optical spectrum has been processed, it is photo-detected and reconverted to the electrical domain where all the aforementioned IMD\textsubscript{3} terms appear, including both direct and crossed contributions, see the RF spectrum shown in Fig. 2b. Note that direct terms lie close to the fundamental tones, while crossed ones depend on the frequency separation between $\omega_{1,2}$ and $\omega_{3,4}$. Calculations yield 2 contributions for the fundamental coefficients and up to 8 contributions for IMD\textsubscript{3} terms (see the supplemental document for detailed derivation). Taking this into account, the current at the PD can be expressed in terms of the fundamental tones $I_1$ and IMD\textsubscript{3} coefficients $I_3$ as
\begin{equation}
\begin{split}
&I_{PD}= R_{PD}|E_{p}|^2\\
&=I_{1,1} \sin\omega_{1,2}t+I_{1,2}\sin\omega_{3,4}t\\
&+I_{3,1}\sin(2\omega_{1,2}-\omega_{2,1})t+I_{3,2}\sin(2\omega_{3,4}-\omega_{4,3})t\\
&+I_{3,3}\sin(2\omega_{1,1,2,2}-\omega_{3,4,3,4})t +I_{3,4}\sin(2\omega_{3,3,4,4}-\omega_{1,2,1,2})t\\
&+I_{3,5}\sin(\omega_{3,3,4,4}-\omega_{4,4,3,3}+\omega_{1,2,1,2})t\\
&+I_{3,6}\sin(\omega_{1,1,2,2}-\omega_{2,2,1,1}+\omega_{3,4,3,4})t\\
&+I_{3,7}\sin(\omega_{1,1}+\omega_{2,2}-\omega_ {3,4})t+I_{3,8}\sin(\omega_{3,3}+\omega_{4,4}-\omega_ {1,2})t,\\
\end{split}
\end{equation}
where $R_{PD}$ is the responsivity of the PD, $I_{3,1}$, $I_{3,2}$ are direct IMD\textsubscript{3} terms represented in red in Fig. 2b, and $I_{3,3}$, $I_{3,4}$, $I_{3,5}$, $I_{3,6}$, $I_{3,7}$ and $I_{3,8}$ are crossed IMD\textsubscript{3} terms, shown in green and purple in Fig. 2b.

By multiplying Eq. (6) by its conjugate we obtain the current at the PD so that we can relate it to all $I_3$ terms using Eq. (7). For the first direct IMD\textsubscript{3} coefficient $I_{3,1}$, this yields
\begin{equation}
\begin{split}
&I_{3,1}= 4 R_{PD}P_i(k-k^2)\cdot \left[  \dfrac{1}{2}\beta \sin \phi_{1} + \Phi_1 A J_{0}^2 J_{1} J_{2} \right] 
\end{split}
\end{equation}
whence $\beta$ is defined as
\begin{equation}
\begin{split}
&\beta=A J_0^2 J_1 J_2 \cos\theta + A J_0 J_1^3 \cos\theta + J_0^2 J_1 J_2,\\
\end{split}
\end{equation}
and $\Phi_1$ as
\begin{equation}
\begin{split}
&\Phi_1=\sin(\phi_{1}-\phi_{2}+\phi_{3}-\theta)+\sin(\phi_{1}+\phi_{3}-\theta)\\
&-\sin(-\phi_{2}+\phi_{3}-\theta)-\sin(\phi_{3}-\theta),
\end{split}
\end{equation}
where $A$ and $\theta$ is the amplitude suppression and phase shift imposed to the OCB by the spectral shaper, respectively. Similarly, the calculation of the second direct IMD\textsubscript{3} coefficient $I_{3,2}$ yields 
\begin{equation}
\begin{split}
&I_{3,2}= 4 R_{PD}P_i(k-k^2)\cdot \left[  \dfrac{1}{2}\beta \sin \phi_{2} - \Phi_2 A J_{0}^2 J_{1} J_{2} \right] 
\end{split}
\end{equation}
where $\Phi_2$ is defined as
\begin{equation}
\begin{split}
&\Phi_2=\sin(\phi_{1}-\phi_{2}+\phi_{3}-\theta)-\sin(\phi_{1}+\phi_{3}+\theta)\\
&+\sin(-\phi_{2}+\phi_{3}+\theta)-\sin(\phi_{3}+\theta).
\end{split}
\end{equation}

On the other hand, assuming $\phi_1=-\phi_2$ and no OCB processing —i.e., $A=1$ and $\theta=0$, all crossed IMD\textsubscript{3} terms ($I_{3,3}$, $I_{3,4}$, $I_{3,5}$, $I_{3,6}$, $I_{3,7}$, $I_{3,8}$) are directly proportional to the phase shifts, and can be expressed by the following expression
\begin{equation}
\begin{split}
I_{3,C} \propto \sin(2\phi_{1}+\phi_{3}) - \sin(\phi_{3}).\\
\end{split}
\end{equation}

To cancel crossed IMD\textsubscript{3} terms, we must first equal Eq. (13) to zero. Fixing $\phi_3=2\phi_1$ the equation is solved for $\phi_3=\pi$, thus $\phi_1=\pi /2 $ and $\phi_2=-\pi /2 $. Now, to calculate $\theta$, we introduce these values in $\Phi_1$ and $\Phi_2$ expressions and equal them to zero as follows
\begin{equation}
  \spalignsys{
  \sin(2\pi-\theta) - \sin(\pi-\theta)=0 ;
  \sin(2\pi+\theta) - \sin(\pi+\theta)=0
  }
\end{equation}
where this system of equations is met when $\theta=\pi$. Finally, to fully cancel direct IMD\textsubscript{3} terms, we solve $\beta$ in Eq. (9) for this value of $\theta$ as 
\begin{equation}
\beta=-A J_0^2 J_1 J_2 - A J_0 J_1^3 + J_0^2 J_1 J_2 = 0,
\end{equation}
applying Taylor series expansion to the third order in $m$ (small signal approximation) we can rewrite this expression as
\begin{equation}
\beta=\dfrac{m^3}{16}(1-3A) = 0,
\end{equation}
where this condition if fulfilled when an amplitude suppression of $A=1/3$ is applied to the OCB, so that direct IMD\textsubscript{3} contributors in Eq. (8) and (10) are theoretically canceled.

On the other hand, calculations (see supplementary for detailed derivation) yield that the coefficients for the fundamental signals $I_{1}$ under small signal condition are
\begin{equation}
\begin{split}
I_{1,1}=-4mA \cos\theta \sin \phi_{1},
\end{split}
\end{equation}
\begin{equation}
\begin{split}
I_{1,2}=-4mA \cos\theta \sin \phi_{2}.
\end{split}
\end{equation}

Note that maximum transmission is achieved because of the quadrature operation of both MZM. In contrast, because of the OCB processing, a penalty of $1/3$ is imposed to the fundamental tones in the optical spectrum, which is translated into MPL losses of 9.4 dB in the RF domain. 

\begin{figure*}[t!]
\centering\includegraphics[scale=0.45]{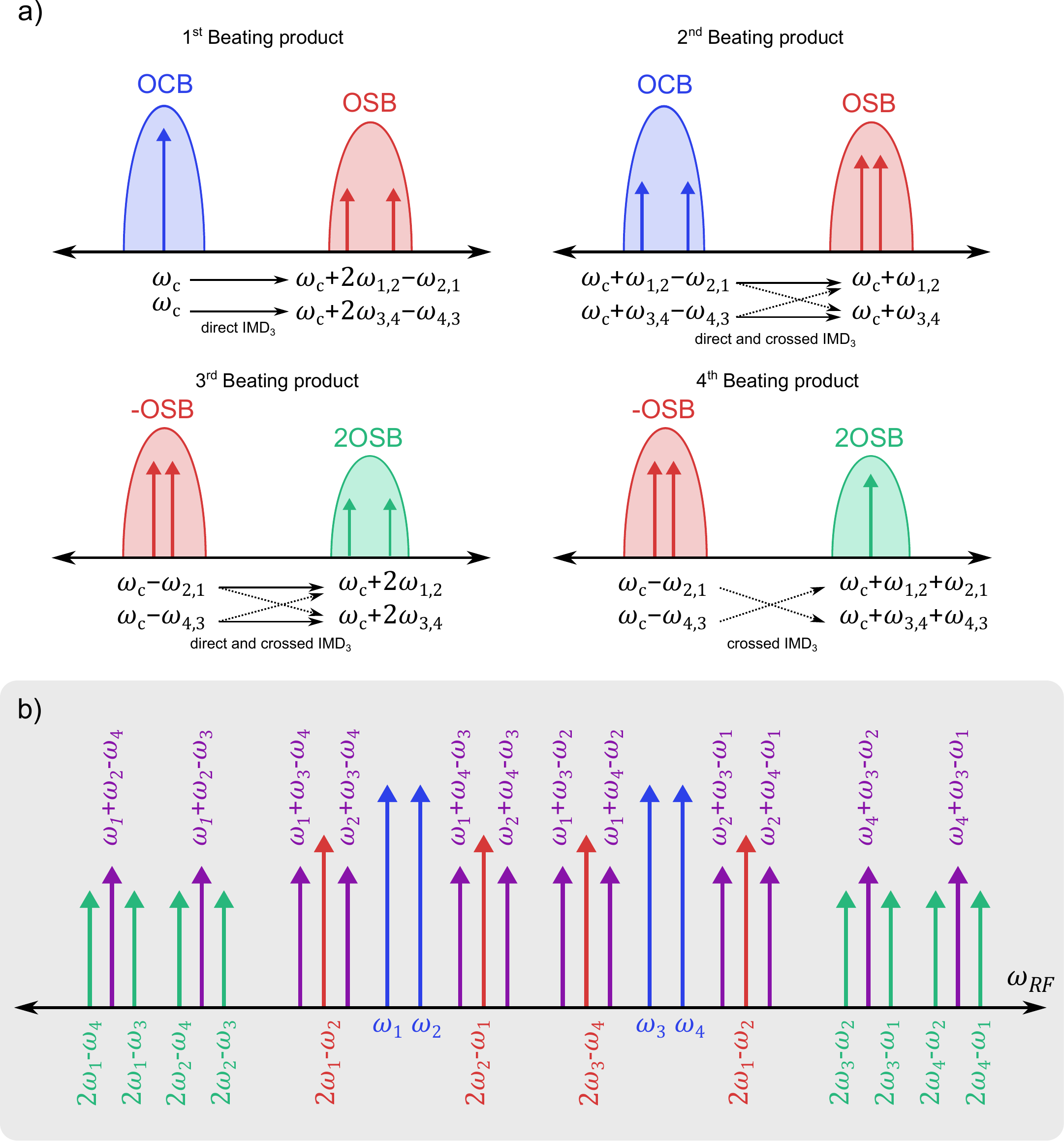}
\caption{a) Representation of the optical modulated spectrum and the different beating products that produce both direct and crossed IMD\textsubscript{3} terms. b) RF spectrum obtained after photodetection considering a four-tone test. In blue the fundamental tones, red direct IMD\textsubscript{3} terms and purple/green crossed IMD\textsubscript{3} terms.}
\end{figure*}

\section{Simulations}

\begin{figure}[b!]
\centering\includegraphics[scale=0.475]{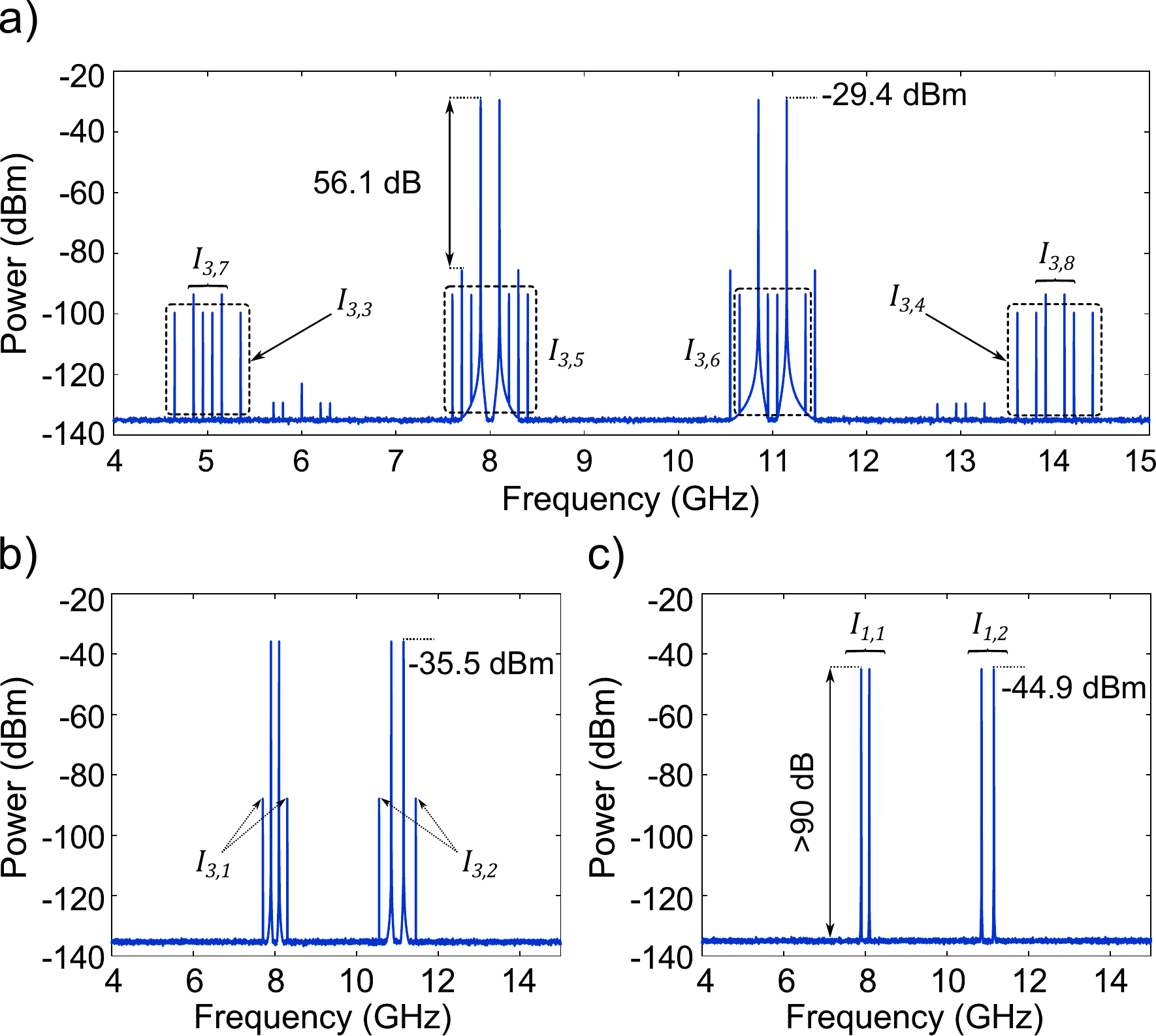}
\caption{a) Simulation of the MPL RF spectrum without linearization ($\phi_1=\pi/2$, $\phi_2=\pi/2$, $\phi_3=0$, $\theta=0$, $A=1$). b) Cancellation of crossed IMD\textsubscript{3} terms ($\phi_1=\pi/2$, $\phi_2=-\pi/2$, $\phi_3=\pi$, $\theta=0$, $A=1$). c) Complete linearization of direct IMD\textsubscript{3} terms with OCB processing ($\phi_1=\pi/2$, $\phi_2=-\pi/2$, $\phi_3=\pi$, $\theta=\pi$, $A=1/3$)}
\end{figure}

To verify the theoretical analysis, several simulations have been carried out using a software tool in Matlab that analytically calculates the electric field output of the DPMZM in the time domain and later computes the fast Fourier transform (FFT) to obtain the resulting spectrum. The input parameters configured in the simulation are: input optical power $P_i=$ 18 dBm, RF power $P_{RF}=$ 8 dBm, $V_{\pi}=$ 5V, $\omega_1=$ 7.9 GHz, $\omega_2=$ 8.1 GHz, $\omega_3=$ 10.85 GHz, $\omega_4=$ 11.15 GHz, $R_{PD}=$ 1 A/W and link losses considered of $L=$ 17 dB. Figure 3a shows the simulation without linearization,  —i.e., with the bias voltages of the DPMZM $\phi_1=\phi_2=\pi/2$, $\phi_3=$ 0 and no OCB processing, $\theta=$ 0 and $A=$ 1. The resulting RF spectrum contains all the IMD\textsubscript{3} theoretically described including both direct and crossed terms. A fundamental to IMD\textsubscript{3} ratio (FIR) of 56.1 dB is obtained at this stage. Figure 3b contains the RF spectrum with only the DC bias voltages adjusted to suppress IMD\textsubscript{3} crossed terms in Eq. (12),  —i.e., $\phi_1=\pi/2$, $\phi_2=-\pi/2$, $\phi_3=\pi$ and no OCB processing, $\theta=0$, $A=1$. Only direct IMD\textsubscript{3} are obtained in the spectrum, as well as the fundamental tones which are reduced 6.1 dB because of the cancellation of the phase relation ($\Phi_{1,2}$) in crossed IMD\textsubscript{3} terms that are also present in $I_1$ coefficients. Figure 3c in turn depicts the linearized RF spectrum with the theoretical values obtained before: $\phi_1=\pi/2$, $\phi_2=-\pi/2$, $\phi_3=\pi$, an attenuation of $A=$ 1/3 and phase of $\theta=\pi$ imposed on the OCB. Simulated results show a FIR over 90 dB, which means an IMD\textsubscript{3} suppression above 33.9 dB compared to Fig. 3a. As this figure illustrates, all IMD\textsubscript{3} terms go below the noise floor and only the fundamental tones are visible in the simulation, which match the theoretical predictions and indicate that the proposed linearization method is feasible to reduce IMD\textsubscript{3}. Note also that the fundamental signals $I_{1,1}$ and $I_{1,2}$ are reduced 9.4 dB compared to the last configuration in Fig. 3b because of the amplitude suppression imposed to the OCB.

\section{Experimental results}

Experimental measurements are carried out following the setup detailed in Fig. 1. A laser source (Tunics T100S) at 1550 nm wavelength and 10 dBm of optical power with a polarization-maintaining fiber is intensity modulated by a DPMZM (Photline MXIQ-LN-40) where up to four tone RF signals are injected. Two of these RF signals centered at 14.99 GHz, separation bandwidth of 3 MHz and power of 6 dBm are generated by a first vector signal generator (E8267C Agilent). Another set of two RF signals centered at 15.01 GHz, separated by 5 MHz and with 6 dBm power are produced by a second vector signal generator (R\&S SMW200A). The modulated signal is thereafter processed by an optical spectral shaper (Waveshaper 4000s), with losses around 10 dB, to impose a certain phase an amplitude on the OCB. To compensate optical losses, an erbium-doped fiber amplifier (EDFA Amonics) is disposed at the output, which is directly connected to the photo-detector (Finisar 70 GHz) to retrieve the electrical signal at the MPL output, measured by a RF spectrum analyzer (R\&S FSW43).

\begin{figure}[b!]
\centering\includegraphics[scale=0.65]{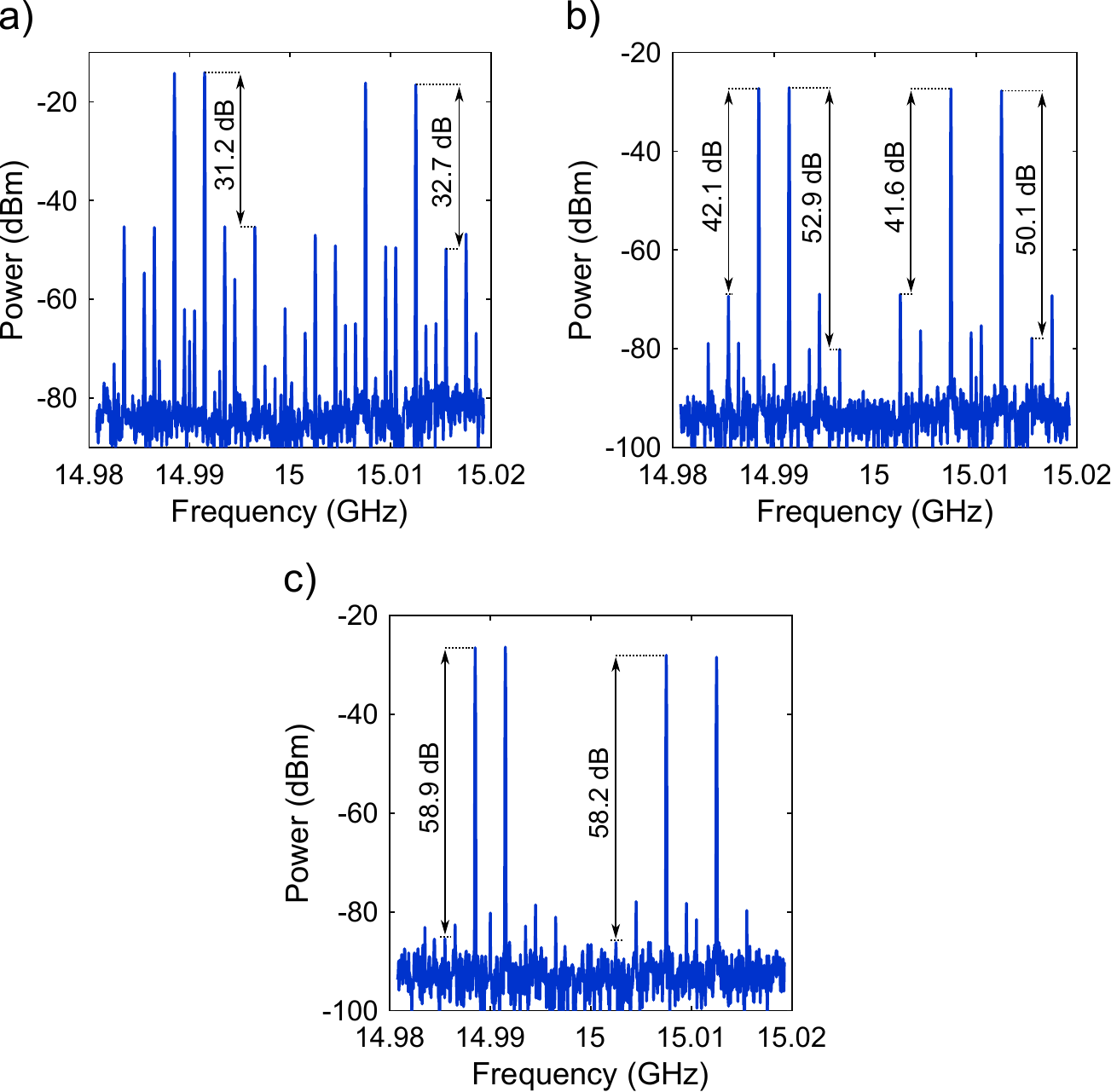}
\caption{Experimental measurements of the retrieved RF spectrum with 10 and 6 dBm of optical and RF power, respectively. a) Reference state without linearization (DC voltages at $\phi_1=\pi/2$, $\phi_2=\pi/2$, $\phi_3=0$) b) linearization of crossed IMD\textsubscript{3} terms ($\phi_1=\pi/2$, $\phi_2=-\pi/2$, $\phi_3=\pi$) and c) linearization of direct IMD\textsubscript{3} terms with OCB processing ($\theta=\pi$ and $A=1/3$).}
\end{figure}

The experimental RF spectrum obtained at the MPL output is shown in Fig. 4 with 10 dBm and 6 dBm of optical and RF power, respectively. Firstly, in Fig. 4a the DPMZM is tuned into a reference state without linearization ($\phi_1=\pi/2$, $\phi_2=\pi/2$, $\phi_3=0$) and no OCB processing, —i.e., is the waveshaper configured as an all-pass filter. At this stage, third-order distortion is clearly visible near the fundamental signals with a FIR  around 32 dB respect to crossed IMD\textsubscript{3} terms. It is worth noting that both RF\textsubscript{1} and RF\textsubscript{2} signals are not completely balanced, as it is seen in the difference between the fundamental coefficients ($I_{1,1}$, $I_{1,2}$). This is caused by fabrication deviations in the DPMZM, which limit its performance and will be present in all experimental measurements. Nevertheless, in Fig. 4b the linearization of crossed IMD\textsubscript{3} terms is measured by properly adjusting the DC bias voltages ($\phi_1=\pi/2$, $\phi_2=-\pi/2$, $\phi_3=\pi$) that have been theoretically calculated. At this stage, the waveshaper continues working as an all-pass filter so no OCB processing is imposed. The FIR of crossed terms is reduced up to 52.9 dB and 50.1 dB, for the case of the fundamental tones centered at 14.99 GHz and 15.01 GHz, respectively. That means a suppression of almost 20 dB for the crossed IMD\textsubscript{3} terms by only tuning the DC voltages, as it was theoretically predicted. However, because of the DPMZM imbalances a degradation around 10 dB is obtained in the fundamental coefficients, which is different from the 6 dB calculated in simulations previously shown. Direct IMD\textsubscript{3} terms at this stage are not canceled and present a FIR around 42 dB. In Fig. 4c, the complete linearized spectrum is shown by processing the OCB with a phase of $\theta=\pi$ and amplitude suppression of $A=1/3$. After applying these filter conditions, a FIR of direct IMD\textsubscript{3} terms of 58.9 dB and 58.2 dB is obtained for the fundamental tones at 14.99 GHz and 15.01 GHz, respectively. That means a suppression of direct terms around 17 dB respect to the previous stage without linearization. The optical power is compensated by the EDFA after applying the filter so that fundamental tones remain constant before and after linearization of direct IMD\textsubscript{3} terms. Specifically, the optical power is increased 9.4 dBm to compensate the amplitude attenuation of 1/3 imposed by the waveshaper. Note also that crossed terms are not fully canceled at this stage and they remain visible above direct terms because of internal deviation errors in the DPMZM. That is why results regarding the SFDR are given respect to the direct IMD\textsubscript{3} terms, which are the critical ones since they are canceled by the OCB processing.

The SFDR measurements are shown in Fig. 5 for both RF\textsubscript{1} and RF\textsubscript{2} fundamental signals. They are studied in separated graphs since they behave differently depending on the filter applied, thus the SDFR obtained might not be the same. RF output power is depicted as a function of increasing RF input powers without and with linearization, —i.e., without and with OCB processing of $\theta=\pi$ and $A=1/3$ with common DC bias voltages at $\phi_1=\pi/2$, $\phi_2=-\pi/2$, $\phi_3=\pi$. A SFDR of 83.17 dB $\cdot$ Hz\textsuperscript{2/3} and 83.98 dB $\cdot$ Hz\textsuperscript{2/3} with a noise floor of -130.79 dBm/Hz is measured without linearization for the case of RF\textsubscript{1} and RF\textsubscript{2} signals, respectively. Likewise, a SFDR of 86.92 dB $\cdot$ Hz\textsuperscript{2/3} and 86.87 dB $\cdot$ Hz\textsuperscript{2/3} with a noise floor of -128.85 dBm/Hz is obtained in turn after linearization for RF\textsubscript{1} and RF\textsubscript{2} signals, respectively. An improvement around 3 dB is observed, which is in part limited by the increase of the noise floor due to the increment of optical power injected by the EDFA at the PD. It is worth noting that the noise floor is relatively high because of the waveshaper and modulator losses, which drastically increases the optical power supplied by the EDFA and so the noise reference level. If we consider the same noise level with and without linearization, a SFDR improvement up to 5 dB could be obtained without compensation penalty.

Compared to other similar approaches, our results in terms of IMD3 suppression and SFDR improvement are very similar to those reported in the linearization of a PM using optical processing \cite{Liu2020} (21.7 dB of IMD3 suppression and a SFDR improvement of 7 dB) and are not as good as those obtained in the linearization of DPMZMs \cite{Jiang2015,Gu2019,Zhu2016} (between 25-45 dB of IMD3 suppression and a SFDR improvement around 11 dB). However, in all these DPMZM examples the linearization has been demonstrated using external RF components and under a two-tone analysis. In contrast, our results are based on optical processing which can be integrated in a photonic chip and the linearization is generalized to simultaneously modulate up to four different signals without IMD3. 

\begin{figure}[t!]
\centering\includegraphics[scale=0.7]{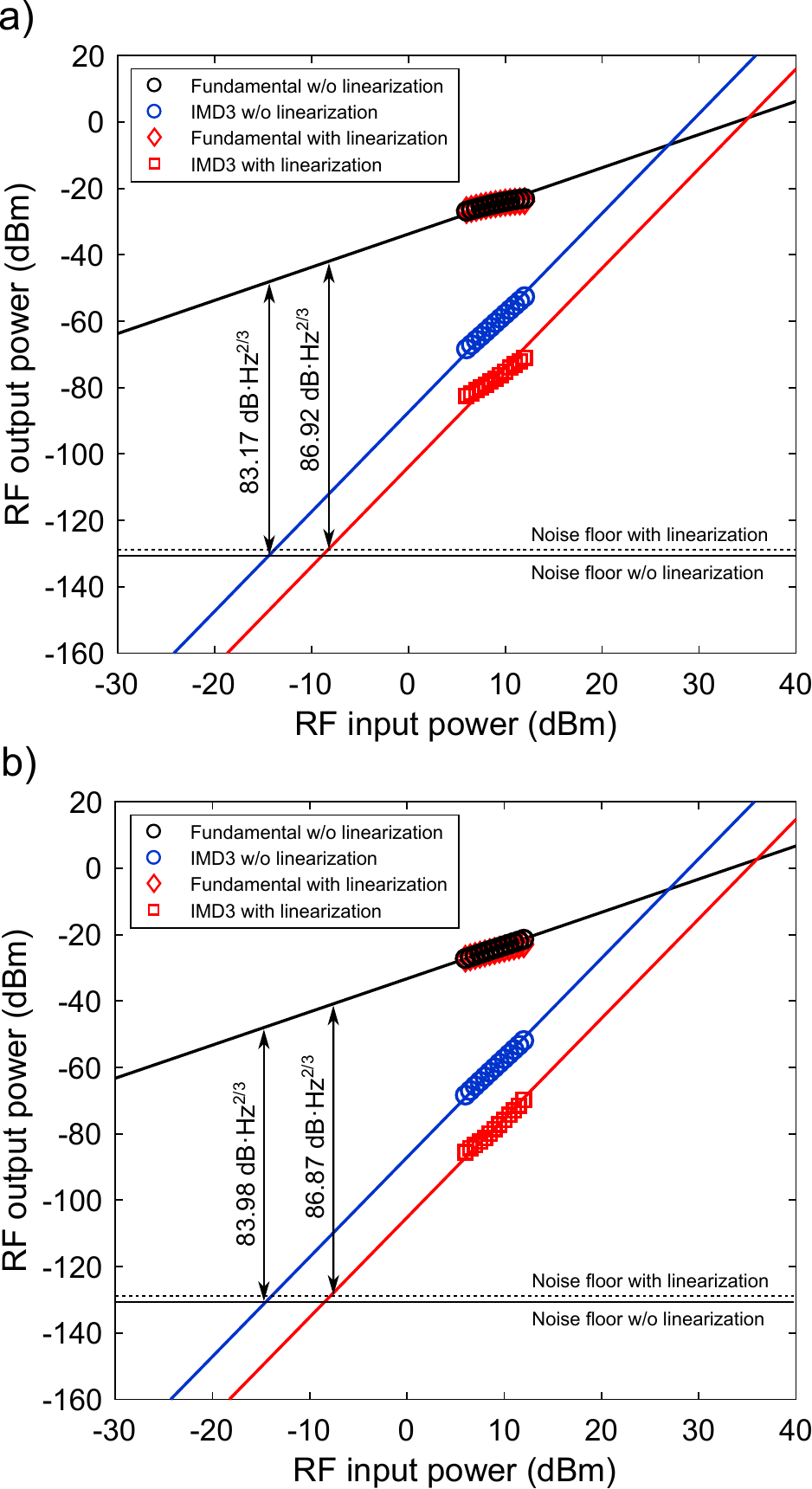}
\caption{Measured SFDR of the MPL before and after linearization (OCB processing of $\theta=\pi$ and $A=1/3$) for a) fundamental tones RF\textsubscript{1} centered at 14.99 GHz and b) RF\textsubscript{2} centered at 15.01 GHz.}
\end{figure}

\section{Discussion}

Although experimental measurements prove that the linearization of a DPMZM by means of OCB processing is feasible, the results shown in Fig. 5 for IMD\textsubscript{3} terms present a slope of 3, instead of 5, which suggest that IMD\textsubscript{3} terms are not completely suppressed. Similar results have been already discussed in other works involving sideband processing \cite{Liu2020}, where it was demonstrated that the non-constant amplitude and phase applied on the modulated spectrum cause incomplete cancellation of IMD\textsubscript{3} terms. In our case, the resolution of the waveshaper (10 GHz) is close to the filter requirements where only the OCB must be processed. That is the reason why relatively high RF frequencies have been configured, mainly to provide a certain bandwidth margin in the modulated optical spectrum where the linearization filter could fit. However, waveshaper filters are designed within the minimum resolution so that the amplitude and phase imposed over the entire OCB bandwidth is not perfectly constant to fully satisfy Eqs. (13) and (15).

To further analyze the behavior of the waveshaper and its influence on the linearization, several deviations in phase and amplitude over the theoretical obtained filter have been experimentally measured. Figure 6 shows the imbalance between fundamental signals ($I_{1,2}-I_{1,1}$) as well as the FIR for RF\textsubscript{1} and RF\textsubscript{2} signals, FIR\textsubscript{1} and FIR\textsubscript{2}, respectively. It is shown in Fig. 6a that OCB phase deviations of $\pm$0.1 over $\pi$ have a  critical influence on the intensity of the fundamental signals in the RF spectrum. For lower values of $\pi$, RF\textsubscript{1} signals centered at 14.99 GHz present higher power values than RF\textsubscript{2} signals at 15.01 GHz, and vice versa, for higher values of $\pi$ RF\textsubscript{2} is higher than RF\textsubscript{1}. Conversely, the amplitude variation has a much lower impact on the imbalance than the phase variation, see Fig. 6b. Note that best results are obtained for a zero deviation in phase where both fundamental signals are balanced ($I_{1,1}=I_{1,2}$), which perfectly match the theoretical predictions.  Figure 6c and d in turn shows the FIR for phase and amplitude deviations. Again, the phase has a higher influence on the results since FIR values drop drastically for deviation errors around $\pm$0.1 over $\pi$. The results demonstrate how critical is the filter design in the linearization method, mainly due to the non-constant phase imposed in the OCB which produces a incomplete suppression of IMD\textsubscript{3} terms. 

\begin{figure}[t!]
\centering\includegraphics[scale=0.475]{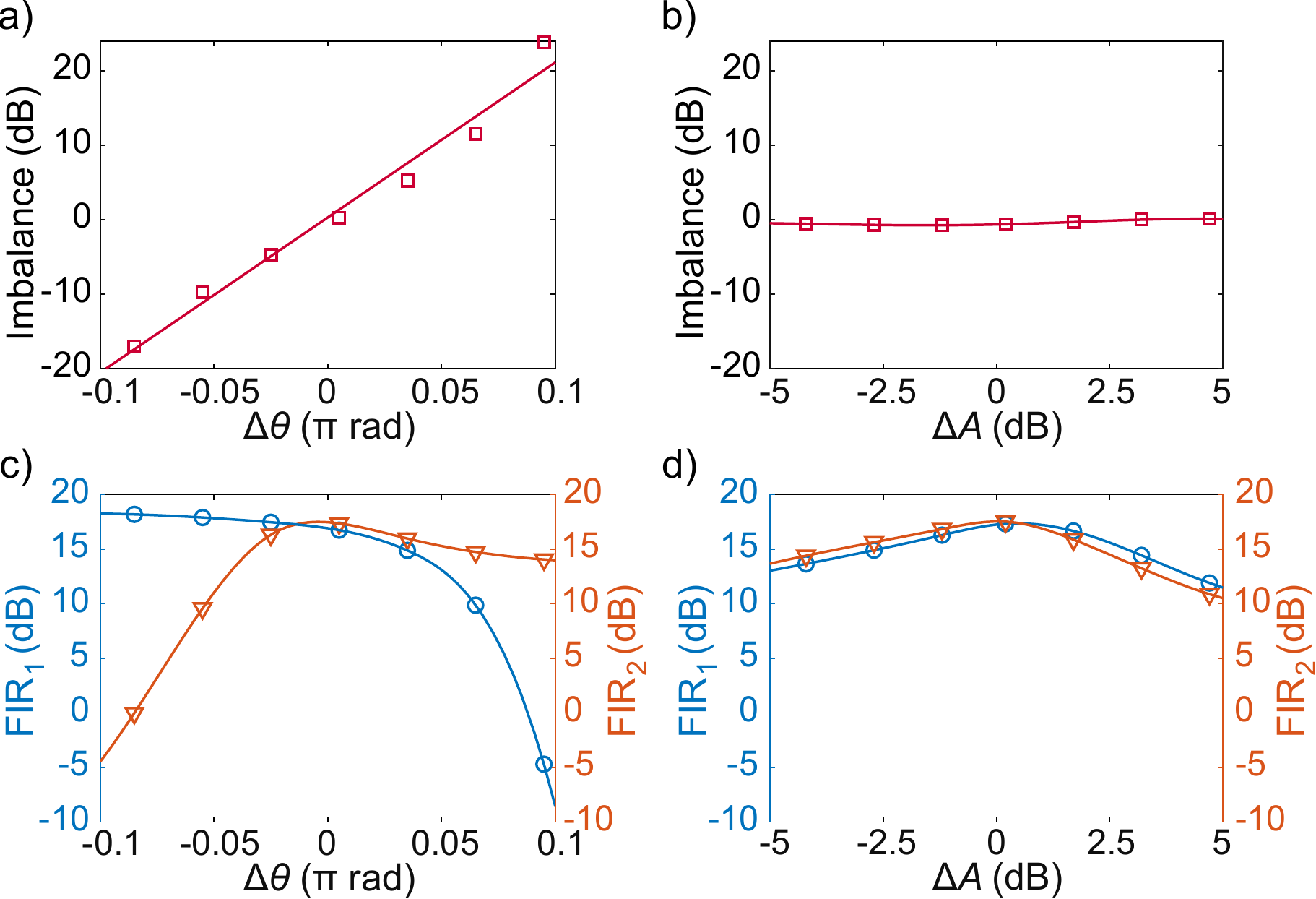}
\caption{Experimental study of the linearization filters. Imbalance between fundamental signals ($I_{1,1}-I_{1,2}$) for a) a certain phase deviation over $\pi$, b) an amplitude deviation over -9.54 dBs. FIR of RF\textsubscript{1} and RF\textsubscript{2} signals for c) the phase and d) amplitude deviation.}
\end{figure}

\section{Conclusion}

In conclusion, OCB processing to linearize a MPL based on a DPMZM is theoretically described and experimentally demonstrated. A four tone test is considered in the study, which allow us to provide an exhaustive mathematical derivation of all IMD\textsubscript{3} terms that are produced in the PD mixing. Simulations show that crossed IMD\textsubscript{3} terms can be canceled out by properly tuning the DC voltages of the modulator, while direct IMD\textsubscript{3} terms need from a precise OCB processing to completely linearize the link. Experimental results are also presented, showing a good agreement with theory and simulations. Specifically, a suppression of crossed and direct IMD\textsubscript{3} terms of 20 dB and 17 dB is measured, and a SFDR improvement of 3 dB is observed in the experiments. Furthermore, a complete study on phase and amplitude deviations in the OCB processing is also carried out, demonstrating how critical is the phase imposed on the MPL performance. Overall, the theoretical description herein presented extends mathematical derivations in previous works to a more complete structure, enabling the modulation of two different RF signals simultaneously, up to 4 frequencies, with complete isolation and no distortion between them. Moreover, the linearization method do not need any extra RF components and other pre-distortion devices, which allows its integration and development in integrated circuits. These results suggest the use of this linearization method in many different MWP applications, such as radio over fiber schemes for high speed and large bandwidth communication systems, among others.

\section*{Acknowledgment}

The authors wish to acknowledge the financial support of Huawei through contract YBN2020095120-SOW3.

\bibliographystyle{IEEEtran}
\bibliography{library}

\newpage


\section*{Supplementary information}

This supplementary material provides a comprehensive mathematical derivation of some of the equations given in the main text about a linearization method for a dual-parallel Mach-Zehnder modulator (DPMZM) using optical side band (OSB) processing. Specifically, we focus on how to calculate the square of the electric field after imposing a certain phase $\theta$ and amplitude $A$ on the OCB, which is related to the current at the photo-detector. By doing this, we obtain all third-order intermodulation distortion (IMD\textsubscript{3}) terms and fundamental $I_1$ coefficients that are studied and analyzed in the main text.

As it is explained in the Eq. (4) of the main text, the electric field after the modulator and its conjugate can be expressed as follows
\begin{equation}
E_{out}(t)\propto e^{j\phi_{3}} \left[ e^{j\phi_{1}} S_1(m) + S_1(-m) \right] + \left[ e^{j\phi_{2}} S_2(m) + S_2(-m) \right],
\end{equation}
\begin{equation}
E^*_{out}(t)\propto e^{-j\phi_{3}} \left[ e^{-j\phi_{1}} S^*_1(m) + S^*_1(-m) \right] + \left[ e^{-j\phi_{2}} S^*_2(m) + S^*_2(-m) \right],
\end{equation}
so that by multiplying them we obtain the square of the field after the OCB processing as
\begin{equation}
\begin{split}
&|E_{p}|^2=E_{p}(t)E^{*}_{p}(t)=\\
& \underbrace{S_1(m) S_1^*(m)}_\text{1} + \underbrace{e^{j\phi_{1}} S_1(m)S_1^*(-m)}_\text{2}+\underbrace{e^{j\phi_{1}}e^{-j\phi_{2}}e^{j\phi_{3}} S_1(m)S_2^*(m)}_\text{3} + \underbrace{e^{j\phi_{1}}e^{j\phi_{3}}S_1(m) S_2^*(-m)}_\text{4} 
\\
& \underbrace{e^{-j\phi_{1}}S_1(-m) S_1^*(m)}_\text{5} + \underbrace{S_1(-m)S_1^*(-m)}_\text{6}+\underbrace{e^{-j\phi_{2}}e^{j\phi_{3}}S_1(-m)S_2^*(m)}_\text{7} + \underbrace{e^{j\phi_{3}}S_1(-m) S_2^*(-m)}_\text{8} 
\\
& \underbrace{e^{-j\phi_{1}}e^{j\phi_{2}}e^{-j\phi_{3}}S_1^*(m) S_2(m)}_\text{9} + \underbrace{e^{j\phi_{2}}e^{-j\phi_{3}}S_1^*(-m)S_2(m)}_\text{10}+\underbrace{S_2(m)S_2^*(m)}_\text{11} + \underbrace{e^{j\phi_{2}}S_2(m) S_2^*(-m)}_\text{12} 
\\
& \underbrace{e^{-j\phi_{1}}e^{-j\phi_{3}}S_1^*(m) S_2(-m)}_\text{13} + \underbrace{e^{-j\phi_{3}}S_1^*(-m)S_2(-m)}_\text{14}+\underbrace{e^{-j\phi_{2}}S_2(-m)S_2^*(m)}_\text{15} + \underbrace{S_2(-m) S_2^*(-m)}_\text{16}.
\end{split}
\end{equation}

Up to 16 contributions are obtained (due to 4x4 multiplication of the field and its conjugate). From these contributions, some of them are direct: $S_1(m) S_1(m)$ or $S_2(m) S_2(m)$ and others crossed $S_1(m) S_2(m)$ or $S_2(m) S_1(m)$. This fact will yields beating products between $\omega_{12}$ and $\omega_{34}$ that will produce crossed $I_3$ terms dependent on $\sin (2\omega_{12}-\omega_{43})t$, for instance. It is also worth noting that all these 16 terms are grouped in pairs, for example, term 1 and 6 are the same but with opposite sign ($m$ is the modulation index and $-m$ refers to the internal push-pull operation of the DPMZM). Specifically these pairs are: 1-6, 2-5, 3-9, 4-13, 7-10, 8-14, 11-16, 12-15. In the next sections, we calculate all these pairs in detail.

\subsection*{Calculation of I\textsubscript{3} coefficients}

Now, we are going to calculate each one of all the contributions shown in Eq. (13), and group them in the aforementioned pairs.

Firstly, let us obtain the first three BP of contribution 1, which is expressed in terms of Bessel functions as
\begin{equation}
\begin{split}
&S_1(m) S_1^*(m)=\\
&\left[ A e^{j\theta}J_{0}^2 + J_{-1}J_{2}(e^{j(2\omega_1-\omega_2)t}+e^{j(2\omega_2-\omega_1)t}) + J_{1}J_{-2}(e^{-j(2\omega_1-\omega_2)t}+e^{-j(2\omega_2-\omega_1)t}) \right]\cdot\\
&\underbrace{\left[ A e^{-j\theta}J_{0}^2 + J_{-1}J_{2}(e^{-j(2\omega_1-\omega_2)t}+e^{-j(2\omega_2-\omega_1)t}) + J_{1}J_{-2}(e^{j(2\omega_1-\omega_2)t}+e^{j(2\omega_2-\omega_1)t}) \right]}_\text{1\textsuperscript{st} beating product}\\
&+\\
&\left[ A e^{j\theta}J_{-1}J_{1}(e^{-j(\omega_1-\omega_2)t}+e^{j(\omega_1-\omega_2)t}) + J_{0}J_{1}(e^{j\omega_1t}+e^{j\omega_2t}) + J_{0}J_{-1}(e^{-j\omega_1t}+e^{-j\omega_2t}) \right]\cdot\\
&\underbrace{\left[ A e^{-j\theta}J_{-1}J_{1}(e^{j(\omega_1-\omega_2)t}+e^{-j(\omega_1-\omega_2)t}) + J_{0}J_{1}(e^{-j\omega_1t}+e^{-j\omega_2t}) + J_{0}J_{-1}(e^{j\omega_1t}+e^{j\omega_2t}) \right]}_\text{2\textsuperscript{nd} beating product}\\
&+\\
&\left[J_{0}J_{1}(e^{j\omega_1t}+e^{j\omega_2t}) + J_{0}J_{-1}(e^{-j\omega_1t}+e^{-j\omega_2t}) +  J_{0}J_{2}(e^{j2\omega_1t}+e^{j2\omega_2t})\right.+\\
&\left. J_{0}J_{-2}(e^{-j2\omega_1t}+e^{-j2\omega_2t}) \right]\cdot\left[J_{0}J_{1}(e^{-j\omega_1t}+e^{-j\omega_2t})\right.+\\
&\underbrace{\left.J_{0}J_{-1}(e^{j\omega_1t}+e^{j\omega_2t})+ J_{0}J_{2}(e^{-j2\omega_1t}+e^{-j2\omega_2t})+ J_{0}J_{-2}(e^{j2\omega_1t}+e^{j2\omega_2t}) \right]}_\text{3\textsuperscript{rd} beating product}\\
&=-4[\underbrace{AJ_{0}^2J_{1}J_{2}\sin\theta}_\text{1\textsuperscript{st} BP}+\underbrace{AJ_{0}J_{1}^3\sin\theta}_\text{2\textsuperscript{nd} BP} + \underbrace{J_{0}^2J_{1}J_{2}}_\text{3\textsuperscript{rd} BP}]\sin(2\omega_{1,2}-\omega_{2,1})t,
\end{split}
\end{equation}
where $A$ and $\theta$ are the amplitude suppression and phase shift imposed to the OCB, respectively.

Now we must calculate other terms considering push-pull configuration, so that $S(-m)$ is introduced, that is $J_{1}(-m)=-J_{1}(m)=J_{-1}(m)$, and therefore $J_{-1}(-m)=-J_{1}(-m)=J_{1}(m)$. Therefore, the calculation of contribution 2, yields
\begin{equation}
\begin{split}
&S_1(m) S_1^*(-m)=\\
&\left[ A e^{j\theta}J_{0}^2 + J_{-1}J_{2}(e^{j(2\omega_1-\omega_2)t}+e^{j(2\omega_2-\omega_1)t}) + J_{1}J_{-2}(e^{-j(2\omega_1-\omega_2)t}+e^{-j(2\omega_2-\omega_1)t}) \right]\cdot\\
&\underbrace{\left[ A e^{-j\theta}J_{0}^2 + J_{1}J_{2}(e^{-j(2\omega_1-\omega_2)t}+e^{-j(2\omega_2-\omega_1)t}) + J_{-1}J_{-2}(e^{j(2\omega_1-\omega_2)t}+e^{j(2\omega_2-\omega_1)t}) \right]}_\text{1\textsuperscript{st} beating product}\\
&+\\
&\left[ A e^{j\theta}J_{-1}J_{1}(e^{-j(\omega_1-\omega_2)t}+e^{j(\omega_1-\omega_2)t}) + J_{0}J_{1}(e^{j\omega_1t}+e^{j\omega_2t}) + J_{0}J_{-1}(e^{-j\omega_1t}+e^{-j\omega_2t}) \right]\cdot\\
&\underbrace{\left[ A e^{-j\theta}J_{1}J_{-1}(e^{j(\omega_1-\omega_2)t}+e^{-j(\omega_1-\omega_2)t}) + J_{0}J_{-1}(e^{-j\omega_1t}+e^{-j\omega_2t}) + J_{0}J_{1}(e^{j\omega_1t}+e^{j\omega_2t}) \right]}_\text{2\textsuperscript{nd} beating product}\\
&+\\
&\left[J_{0}J_{1}(e^{j\omega_1t}+e^{j\omega_2t}) + J_{0}J_{-1}(e^{-j\omega_1t}+e^{-j\omega_2t}) + J_{0}J_{2}(e^{j2\omega_1t}+e^{j2\omega_2t})\right.+\\
&\left. J_{0}J_{-2}(e^{-j2\omega_1t}+e^{-j2\omega_2t}) \right]\cdot\left[J_{0}J_{-1}(e^{-j\omega_1t}+e^{-j\omega_2t})\right.+\\
&\underbrace{\left.J_{0}J_{1}(e^{j\omega_1t}+e^{j\omega_2t})+J_{0}J_{2}(e^{-j2\omega_1t}+e^{-j2\omega_2t})+J_{0}J_{-2}(e^{j2\omega_1t}+e^{j2\omega_2t}) \right]}_\text{3\textsuperscript{rd} beating product}\\
&\\
&=-4[\underbrace{AJ_{0}^2J_{1}J_{2}\cos\theta}_\text{1\textsuperscript{st} BP}+\underbrace{AJ_{0}J_{1}^3\cos\theta}_\text{2\textsuperscript{nd} BP} + \underbrace{J_{0}^2J_{1}J_{2}}_\text{3\textsuperscript{rd} BP}]\sin(2\omega_{1,2}-\omega_{2,1})t.
\end{split}
\end{equation}

For the sake of simplicity, we can define two new variables with the results in brackets of Eq. (S4) and (S5) since they will appear in other calculations:
\begin{equation}
\alpha=A J_0^2 J_1 J_2 \sin\theta + A J_0 J_1^3 \sin\theta + J_0^2 J_1 J_2
\end{equation}
\begin{equation}
\beta=A J_0^2 J_1 J_2 \cos\theta + A J_0 J_1^3 \cos\theta + J_0^2 J_1 J_2
\end{equation}

Having this in mind, let us begin with pairs including direct contributions (1-6, 2-5, 11-16, 12-15), which are easier to calculate. For the first case, the calculation \textbf{of pair 1-6} yields
\begin{equation}
S_1(m)S^*_1(m) + S_1(-m)S^*_1(-m) = -\alpha + \alpha = 0,
\end{equation}
second \textbf{pair 2-5} can be expressed as
\begin{equation}
\begin{split}
&e^{j\phi_{1}}S_1(m)S^*_1(-m) + e^{-j\phi_{1}}S_1(-m)S^*_1(m) =
\\
&-j\beta e^{j\phi_{1}} + j\beta e^{-j\phi_{1}} = j\beta(-e^{j\phi_{1}}+e^{-j\phi_{1}}) = 2\beta \sin \phi_{1} \sin(2\omega_{1,2}-\omega_{2,1}),
\end{split}
\end{equation}
\textbf{pair 11-16} can be calculated as follows
\begin{equation}
S_2(m)S^*_2(m) + S_2(-m)S^*_2(-m) = -\alpha + \alpha = 0,
\end{equation}
and finally last direct contribution, \textbf{pair 12-15} is expressed as
\begin{equation}
\begin{split}
&e^{j\phi_{2}}S_2(m)S^*_2(-m) + e^{-j\phi_{2}}S_2(-m)S^*_2(m) =
\\
&-j\beta e^{j\phi_{2}} + j\beta e^{-j\phi_{2}} = j\beta(-e^{j\phi_{2}}+e^{-j\phi_{2}}) = 2\beta \sin \phi_{2} \sin(2\omega_{3,4}-\omega_{4,3}).
\end{split}
\end{equation}

So far, all the considered contributions only produce \textbf{direct IMD\textsubscript{3} terms}, either for RF1 ($\omega_{1,2}$) or RF2 ($\omega_{3,4}$).

Conversely, now we analyze pairs that include crossed contributions so that four BPs are considered. These are 3-9, 4-13, 7-10, 8-14 and they can produce both \textbf{direct and crossed IMD\textsubscript{3} terms}.

Let us begin with \textbf{pair 3-9} which can be calculated as follows
\begin{equation}
\begin{split}
&e^{j\phi_{1}}e^{-j\phi_{2}}e^{j\phi_{3}} S_1(m)S_2^*(m) + e^{-j\phi_{1}}e^{j\phi_{2}}e^{-j\phi_{3}}S_1^*(m) S_2(m)=\\
\end{split}
\end{equation}
\begin{align*}
&\underbrace{-4A\sin(\chi+\theta)  J_{0}^2 J_{1} J_{2} \sin(2\omega_{3,4}-\omega_{4,3})t+4A\sin(\chi-\theta)  J_{0}^2 J_{1} J_{2} \sin(2\omega_{1,2}-\omega_{2,1})t}_\text{1\textsuperscript{st} beating product}+\\
&-4A\sin(\chi +\theta)J_{0} J_{1}^3[\sin(\omega_1-\omega_2+\omega_3)t]-4A\sin(\chi+\theta)J_{0} J_{1}^3[\sin(\omega_1-\omega_2+\omega_4)t]\\
&-4A\sin(\chi+\theta)J_{0} J_{1}^3[\sin(\omega_2-\omega_1+\omega_3)t] -4A\sin(\chi+\theta)J_{0} J_{1}^3[\sin(\omega_2-\omega_1+\omega_4)t]\\
&+4A\sin(\chi-\theta)J_{0} J_{1}^3[\sin(\omega_3-\omega_4+\omega_1)t]+4A\sin(\chi -\theta)J_{0} J_{1}^3[\sin(\omega_3-\omega_4+\omega_2)t]\\
&\underbrace{+4A\sin(\chi-\theta)J_{0} J_{1}^3[\sin(\omega_4-\omega_3+\omega_1)t]+4A\sin(\chi-\theta)J_{0} J_{1}^3[\sin(\omega_4-\omega_3+\omega_2)t]}_\text{2\textsuperscript{nd} beating product}+\\
&- 4 \sin(\chi) J_{0}^2J_{1}J_{2}\sin(2\omega_1-\omega_3)t- 4 \sin(\chi) J_{0}^2J_{1}J_{2}\sin(2\omega_1-\omega_4)t\\
&- 4 \sin(\chi) J_{0}^2J_{1}J_{2}\sin(2\omega_2-\omega_3)t- 4 \sin(\chi) J_{0}^2J_{1}J_{2}\sin(2\omega_2-\omega_4)t\\
&+ 4 \sin(\chi) J_{0}^2J_{1}J_{2}\sin(2\omega_3-\omega_1)t+ 4 \sin(\chi) J_{0}^2J_{1}J_{2}\sin(2\omega_3-\omega_2)t\\
&\underbrace{+ 4 \sin(\chi) J_{0}^2J_{1}J_{2}\sin(2\omega_4-\omega_1)t+ 4 \sin(\chi) TJ_{0}^2J_{1}J_{2}\sin(2\omega_4-\omega_2)t}_\text{3\textsuperscript{rd} beating product}+\\
&-4\sin(\chi)J_{0} J_{1}^3[\sin(\omega_1+\omega_2-\omega_3)t]-4\sin(\chi)J_{0} J_{1}^3[\sin(\omega_1+\omega_2-\omega_4)t]\\
&\underbrace{+4\sin(\chi)J_{0} J_{1}^3[\sin(\omega_3+\omega_4-\omega_1)t]+4\sin(\chi)J_{0} J_{1}^3[\sin(\omega_3+\omega_4-\omega_2)t]}_\text{4\textsuperscript{th} beating product}.
\end{align*}
where $\chi$ is  $\phi_{1}-\phi_{2}+\phi_{3}$. Note that now different direct and crossed IMD\textsubscript{3} terms are obtained. Later, we will group each one in I\textsubscript{3} coefficients.

\textbf{Pair 4-13} in turn can be calculated as
\begin{equation}
\begin{split}
&e^{j\phi_{1}}e^{j\phi_{3}}S_1(m) S_2^*(-m)+e^{-j\phi_{1}}e^{-j\phi_{3}}S_1^*(m) S_2(-m)=\\
\end{split}
\end{equation}
\begin{align*}
&\underbrace{+4 A \sin(\phi_{1}+\phi_{3}+\theta) J_{0}^2J_{1}J_{2}   \sin(2\omega_{3,4}-\omega_{4,3})t+4 A \sin(\phi_{1}+\phi_{3}-\theta) J_{0}^2J_{1}J_{2} \sin(2\omega_{1,2}-\omega_{2,1})t+}_\text{1\textsuperscript{st} beating product}\\
&+4A\sin(\phi_{1}+\phi_{3}+\theta)J_{0} J_{1}^3[\sin(\omega_1-\omega_2+\omega_3)t]+4A\sin(\phi_{1}+\phi_{3}+\theta)J_{0} J_{1}^3[\sin(\omega_1-\omega_2+\omega_4)t]\\
&+4A\sin(\phi_{1}+\phi_{3}+\theta)J_{0} J_{1}^3[\sin(\omega_2-\omega_1+\omega_3)t]+4A\sin(\phi_{1}+\phi_{3}+\theta)J_{0} J_{1}^3[\sin(\omega_2-\omega_1+\omega_4)t]\\
&+4A\sin(\phi_{1}+\phi_{3}-\theta)J_{0} J_{1}^3[\sin(\omega_3-\omega_4+\omega_1)t]+4A\sin(\phi_{1}+\phi_{3}-\theta)J_{0} J_{1}^3[\sin(\omega_3-\omega_4+\omega_2)t]\\
&\underbrace{+4A\sin(\phi_{1}+\phi_{3}-\theta)J_{0} J_{1}^3[\sin(\omega_4-\omega_3+\omega_1)t]+4A\sin(\phi_{1}+\phi_{3}-\theta)J_{0} J_{1}^3[\sin(\omega_4-\omega_3+\omega_2)t]}_\text{2\textsuperscript{nd} beating product}+\\
&+ 4 \sin(\phi_{1}+\phi_{3}) J_{0}^2J_{1}J_{2}  \sin(2\omega_1-\omega_3)t+ 4  \sin(\phi_{1}+\phi_{3}) J_{0}^2J_{1}J_{2} \sin(2\omega_1-\omega_4)t\\
&+ 4 \sin(\phi_{1}+\phi_{3}) J_{0}^2J_{1}J_{2}  \sin(2\omega_2-\omega_3)t+ 4 \sin(\phi_{1}+\phi_{3}) J_{0}^2J_{1}J_{2}  \sin(2\omega_2-\omega_4)t\\
&+ 4 \sin(\phi_{1}+\phi_{3}) J_{0}^2J_{1}J_{2}  \sin(2\omega_3-\omega_1)t+ 4 \sin(\phi_{1}+\phi_{3}) J_{0}^2J_{1}J_{2}  \sin(2\omega_3-\omega_2)t\\
&\underbrace{+ 4 \sin(\phi_{1}+\phi_{3}) J_{0}^2J_{1}J_{2}  \sin(2\omega_4-\omega_1)t+ 4 \sin(\phi_{1}+\phi_{3}) J_{0}^2J_{1}J_{2}  \sin(2\omega_4-\omega_2)t}_\text{3\textsuperscript{rd} beating product}+\\
&+4\sin(\phi_{1}+\phi_{3})J_{0} J_{1}^3[\sin(\omega_1+\omega_2-\omega_3)t]+4\sin(\phi_{1}+\phi_{3})J_{0} J_{1}^3[\sin(\omega_1+\omega_2-\omega_4)t]\\
&\underbrace{+4\sin(\phi_{1}+\phi_{3})J_{0} J_{1}^3[\sin(\omega_3+\omega_4-\omega_1)t]+4\sin(\phi_{1}+\phi_{3})J_{0} J_{1}^3[\sin(\omega_3+\omega_4-\omega_2)t]}_\text{4\textsuperscript{th} beating product}.
\end{align*}

\textbf{Pair 7-10} calculation yields
\begin{equation}
\begin{split}
&e^{-j\phi_{2}}e^{j\phi_{3}}S_1(-m) S_2^*(m)+e^{j\phi_{2}}e^{-j\phi_{3}}S_1^*(-m) S_2(m)=\\
\end{split}
\end{equation}
\begin{align*}
&\underbrace{-4A \sin(-\phi_{2}+\phi_{3}+\theta) J_{0}^2J_{1}J_{2}   \sin(2\omega_{3,4}-\omega_{4,3})t -4A \sin(-\phi_{2}+\phi_{3}-\theta) J_{0}^2J_{1}J_{2}   \sin(2\omega_{1,2}-\omega_{2,1})t}_\text{1\textsuperscript{st} beating product}+\\
&-4A\sin(-\phi_{2}+\phi_{3}+\theta)J_{0} J_{1}^3[\sin(\omega_1-\omega_2+\omega_3)t]-4A\sin(-\phi_{2}+\phi_{3}+\theta)J_{0} J_{1}^3[\sin(\omega_1-\omega_2+\omega_4)t]\\
&-4A\sin(-\phi_{2}+\phi_{3}+\theta)J_{0} J_{1}^3[\sin(\omega_2-\omega_1+\omega_3)t]-4A\sin(-\phi_{2}+\phi_{3}+\theta)J_{0} J_{1}^3[\sin(\omega_2-\omega_1+\omega_4)t]\\
&-4A\sin(-\phi_{2}+\phi_{3}-\theta)J_{0} J_{1}^3[\sin(\omega_3-\omega_4+\omega_1)t]-4A\sin(-\phi_{2}+\phi_{3}-\theta)J_{0} J_{1}^3[\sin(\omega_3-\omega_4+\omega_2)t]\\
&\underbrace{-4A\sin(-\phi_{2}+\phi_{3}-\theta)J_{0} J_{1}^3[\sin(\omega_4-\omega_3+\omega_1)t]-4A\sin(-\phi_{2}+\phi_{3}-\theta)J_{0} J_{1}^3[\sin(\omega_4-\omega_3+\omega_2)t]}_\text{2\textsuperscript{nd} beating product}+\\
&- 4 J_{0}^2J_{1}J_{2} \sin(-\phi_{2}+\phi_{3}) \sin(2\omega_1-\omega_3)t- 4 J_{0}^2J_{1}J_{2} \sin(-\phi_{2}+\phi_{3}) \sin(2\omega_1-\omega_4)t\\
&- 4 J_{0}^2J_{1}J_{2} \sin(-\phi_{2}+\phi_{3}) \sin(2\omega_2-\omega_3)t- 4 J_{0}^2J_{1}J_{2} \sin(-\phi_{2}+\phi_{3}) \sin(2\omega_2-\omega_4)t\\
&- 4 J_{0}^2J_{1}J_{2} \sin(-\phi_{2}+\phi_{3}) \sin(2\omega_3-\omega_1)t- 4 J_{0}^2J_{1}J_{2} \sin(-\phi_{2}+\phi_{3}) \sin(2\omega_3-\omega_2)t\\
&\underbrace{- 4 J_{0}^2J_{1}J_{2} \sin(-\phi_{2}+\phi_{3}) \sin(2\omega_4-\omega_1)t- 4 J_{0}^2J_{1}J_{2} \sin(-\phi_{2}+\phi_{3}) \sin(2\omega_4-\omega_2)t}_\text{3\textsuperscript{rd} beating product}+\\
&-4\sin(-\phi_{2}+\phi_{3})J_{0} J_{1}^3[\sin(\omega_1+\omega_2-\omega_3)t]-4\sin(-\phi_{2}+\phi_{3})J_{0} J_{1}^3[\sin(\omega_1+\omega_2-\omega_4)t]\\
&\underbrace{-4\sin(-\phi_{2}+\phi_{3})J_{0} J_{1}^3[\sin(\omega_3+\omega_4-\omega_1)t]-4\sin(-\phi_{2}+\phi_{3})J_{0} J_{1}^3[\sin(\omega_3+\omega_4-\omega_2)t]}_\text{4\textsuperscript{th} beating product}.
\end{align*}
\newpage

\textbf{Pair 8-14} can be calculated as
\begin{equation}
\begin{split}
& e^{j\phi_{3}}S_1(-m) S_2^*(-m) + e^{-j\phi_{3}}S_1^*(-m)S_2(-m)=\\
\end{split}
\end{equation}
\begin{align*}
& \underbrace{4A \sin(\phi_{3}+\theta) J_{0}^2J_{1}J_{2}  \sin(2\omega_{3,4}-\omega_{4,3})t -4A \sin(\phi_{3}-\theta) J_{0}^2J_{1}J_{2}  \sin(2\omega_{1,2}-\omega_{2,1})t}_\text{1\textsuperscript{st} beating product}+\\
&+4A\sin(\phi_{3}+\theta)J_{0} J_{1}^3[\sin(\omega_1-\omega_2+\omega_3)t]+4A\sin(\phi_{3}+\theta)J_{0} J_{1}^3[\sin(\omega_1-\omega_2+\omega_4)t]\\
&+4A\sin(\phi_{3}+\theta)J_{0} J_{1}^3[\sin(\omega_2-\omega_1+\omega_3)t]+4A\sin(\phi_{3}+\theta)J_{0} J_{1}^3[\sin(\omega_2-\omega_1+\omega_4)t]\\
&-4A\sin(\phi_{3}-\theta)J_{0} J_{1}^3[\sin(\omega_3-\omega_4+\omega_1)t]-4A\sin(\phi_{3}-\theta)J_{0} J_{1}^3[\sin(\omega_3-\omega_4+\omega_2)t]\\
&\underbrace{-4A\sin(\phi_{3}-\theta)J_{0} J_{1}^3[\sin(\omega_4-\omega_3+\omega_1)t]-4A\sin(\phi_{3}-\theta)J_{0} J_{1}^3[\sin(\omega_4-\omega_3+\omega_2)t]}_\text{2\textsuperscript{nd} beating product}+\\
&+ 4  J_{0}^2J_{1}J_{2}  \sin(\phi_{3}) \sin(2\omega_1-\omega_3)t+ 4  J_{0}^2J_{1}J_{2}  \sin(\phi_{3}) \sin(2\omega_1-\omega_4)t\\
&+ 4  J_{0}^2J_{1}J_{2}  \sin(\phi_{3}) \sin(2\omega_2-\omega_3)t+ 4  J_{0}^2J_{1}J_{2}  \sin(\phi_{3}) \sin(2\omega_2-\omega_4)t\\
&- 4  J_{0}^2J_{1}J_{2}  \sin(\phi_{3}) \sin(2\omega_3-\omega_1)t- 4  J_{0}^2J_{1}J_{2}  \sin(\phi_{3}) \sin(2\omega_3-\omega_2)t\\
&\underbrace{- 4  J_{0}^2J_{1}J_{2}  \sin(\phi_{3}) \sin(2\omega_4-\omega_1)t- 4  J_{0}^2J_{1}J_{2}  \sin(\phi_{3}) \sin(2\omega_4-\omega_2)t}_\text{3\textsuperscript{rd} beating product}+\\
&+4\sin(\phi_{3})J_{0} J_{1}^3[\sin(\omega_1+\omega_2-\omega_3)t]+4\sin(\phi_{3})J_{0} J_{1}^3[\sin(\omega_1+\omega_2-\omega_4)t]\\
&\underbrace{-4\sin(\phi_{3})J_{0} J_{1}^3[\sin(\omega_3+\omega_4-\omega_1)t]-4\sin(\phi_{3})J_{0} J_{1}^3[\sin(\omega_3+\omega_4-\omega_2)t]}_\text{4\textsuperscript{th} beating product}.
\end{align*}

To calculate direct IMD\textsubscript{3} terms, we can relate Eqs. (S8, S9, S10, S11, S12, S13, S14, S15) with each coefficients using the following expression
\begin{equation}
\begin{split}
& I_{3,1}\sin(2\omega_{1,2}-\omega_{2,1}),\\
& I_{3,2}\sin(2\omega_{3,4}-\omega_{4,3}),
\end{split}
\end{equation}
yielding for $I_{3,1}$:
\begin{equation}
\begin{split}
 I_{3,1}= &R_{PD}P_i(k-k^2)[2\beta \sin \phi_{1} + 4A J_{0}^2 J_{1} J_{2} [\sin(\phi_{1}-\phi_{2}+\phi_{3}-\theta)+\\
&\sin(\phi_{1}+\phi_{3}-\theta) -\sin(-\phi_{2}+\phi_{3}-\theta)-\sin(\phi_{3}-\theta)]],\\
\end{split}
\end{equation}
whence $\beta$ is defined as
\begin{equation}
\beta=A J_0^2 J_1 J_2 \cos\theta + A J_0 J_1^3 \cos\theta + J_0^2 J_1 J_2,
\end{equation}
and for $I_{3,2}$:
\begin{equation}
\begin{split}
I_{3,2}= &R_{PD}P_i(k-k^2)[2\beta \sin \phi_{2} - 4A J_{0}^2 J_{1} J_{2} [\sin(\phi_{1}-\phi_{2}+\phi_{3}+\theta)-\\
&\sin(\phi_{1}+\phi_{3}+\theta) +\sin(-\phi_{2}+\phi_{3}+\theta)-\sin(\phi_{3}+\theta)]],\\
\end{split}
\end{equation}
so that we obtain the same results given in Eq. (8) and (10) of the main text.

Likewise, to calculate crossed IMD\textsubscript{3} terms we relate same Eqs. (S8, S9, S10, S11, S12, S13, S14, S15) with each coefficient using the next expression
\begin{equation}
\begin{split}
&I_{3,3}\sin(2\omega_{1,1,2,2}-\omega_{3,4,3,4})t,\\
&I_{3,4}\sin(2\omega_{3,3,4,4}-\omega_{1,2,1,2})t,\\
&I_{3,5}\sin(\omega_{1,1,2,2}-\omega_{2,2,1,1}+\omega_{3,4,3,4})t,\\
&I_{3,6}\sin(\omega_{3,3,4,4}-\omega_{4,4,3,3}+\omega_{1,2,1,2})t,\\
&I_{3,7}\sin(\omega_{1,1}+\omega_{2,2}-\omega_{3,4})t,\\
&I_{3,8}\sin(\omega_{1,1}+\omega_{2,2}-\omega_{3,4})t.
\end{split}
\end{equation}
which yields for $I_{3,3}$:
\begin{equation}
\begin{split}
&I_{3,3}=- 4R_{PD}P_i(k-k^2) J_{0}^2J_{1}J_{2} \left[  \sin(\phi_{1}-\phi_{2}+\phi_{3}) - \sin(\phi_{1}+\phi_{3}) + \sin(-\phi_{2}+\phi_{3}) - \sin(\phi_{3})\right] ,
\end{split}
\end{equation}
for $I_{3,4}$:
\begin{equation}
\begin{split}
&I_{3,4}= 4R_{PD}P_i(k-k^2) J_{0}^2J_{1}J_{2}  \left[ \sin(\phi_{1}-\phi_{2}+\phi_{3})  + \sin(\phi_{1}+\phi_{3})  - \sin(-\phi_{2}+\phi_{3})  - \sin(\phi_{3}) \right] ,
\end{split}
\end{equation}
for $I_{3,5}$:
\begin{equation}
\begin{split}
I_{3,5}=&-4R_{PD}P_i(k-k^2)AJ_0J_1^3 [ \sin(\phi_{1}-\phi_{2}+\phi_{3}+\theta)-\sin(\phi_{1}+\phi_{3}+\theta)\\
&+\sin(-\phi_{2}+\phi_{3}+\theta)-\sin(\phi_{3}+\theta) ] ,
\end{split}
\end{equation}
for $I_{3,6}$:
\begin{equation}
\begin{split}
I_{3,6}=&4R_{PD}P_i(k-k^2)AJ_0J_1^3 [ \sin(\phi_{1}-\phi_{2}+\phi_{3}-\theta)+\sin(\phi_{1}+\phi_{3}-\theta)-\\
&\sin(-\phi_{2}+\phi_{3}-\theta)-\sin(\phi_{3}-\theta) ] ,
\end{split}
\end{equation}
for $I_{3,7}$:
\begin{equation}
\begin{split}
&I_{3,7}=-4R_{PD}P_i(k-k^2)J_0J_1^3 \left[ \sin(\phi_{1}-\phi_{2}+\phi_{3})-\sin(\phi_{1}+\phi_{3})+\sin(-\phi_{2}+\phi_{3})-\sin(\phi_{3}) \right] ,
\end{split}
\end{equation}
and for $I_{3,8}$:
\begin{equation}
\begin{split}
&I_{3,8}=4R_{PD}P_i(k-k^2)J_0J_1^3 \left[ \sin(\phi_{1}-\phi_{2}+\phi_{3})+\sin(\phi_{1}+\phi_{3})-\sin(-\phi_{2}+\phi_{3})-\sin(\phi_{3}) \right] .
\end{split}
\end{equation}

Assuming $\phi_{1}=-\phi_{2}$, $A=1$ and $\theta=0$, all crossed IMD\textsubscript{3} terms are proportional to the phase relation given in the following expression
\begin{equation}
\begin{split}
I_{3,C} \propto \sin(2\phi_{1}+\phi_{3}) - \sin(\phi_{3}),\\
\end{split}
\end{equation}
which is the Eq. (11) of the main text.

\subsection*{Calculation of I\textsubscript{1} coefficients}

The main BP that produces the fundamental coefficients is the mixing between $\omega_c$ and $\omega_c+\omega_{1,2} (\omega_c-\omega_{1,2})$. Calculations of this BP for $I_{1,1}$ yield
\begin{equation}
\begin{split}
I_{1,1}=&-4R_{PD}P_i(k-k^2)\sin\omega_{1,2}t[2A J_0^3 J_1 \cos\theta \sin \phi_{1}\\
&+AJ_{0}^3J_{1}\sin(\phi_{1}-\phi_{2}+\phi_{3}-\theta)+AJ_{0}^3J_{1}\sin(\phi_{1}+\phi_{3}-\theta)\\
&-AJ_{0}^3J_{1}\sin(-\phi_{2}+\phi_{3}-\theta)-AJ_{0}^3J_{1}\sin(\phi_{3}-\theta)]\\
\end{split}
\end{equation}

Given that $\phi_{1}=\pi/2$, $\phi_{2}=-\pi/2$, $\phi_{3}=\theta=\pi$, and $A=1/3$ we obtain
\begin{equation}
\begin{split}
&I_{1,1} =-8A J_0^3 J_1 \cos\theta \sin \phi_{1}\sin\omega_{1,2}t,
\end{split}
\end{equation}
and under small signal condition we get
\begin{equation}
\begin{split}
&I_{1,1}=-4mA \cos\theta \sin \phi_{1}=-\dfrac{4}{3}m.
\end{split}
\end{equation}

Similarly, for $I_{1,2}$ we get 
\begin{equation}
\begin{split}
I_{1,2} =&-4R_{PD}P_i(k-k^2)\sin\omega_{3,4}t[2A J_0^3 J_1 \cos\theta \sin \phi_{2}\\
&-AJ_{0}^3J_{1}\sin(\phi_{1}-\phi_{2}+\phi_{3}+\theta)+AJ_{0}^3J_{1}\sin(\phi_{1}+\phi_{3}+\theta)\\
&-AJ_{0}^3J_{1}\sin(-\phi_{2}+\phi_{3}+\theta)+AJ_{0}^3J_{1}\sin(\phi_{3}+\theta)]\\
\end{split}
\end{equation}

Substituting we obtain
\begin{equation}
\begin{split}
&I_{1,2} \sin\omega_{3,4} =-8a J_0^3 J_1 \cos\theta \sin \phi_{2}\sin\omega_{3,4}t,
\end{split}
\end{equation}
under small signal condition we get
\begin{equation}
\begin{split}
&I_{1,2}=-4ma \cos\theta \sin \phi_{2}=\dfrac{4}{3}m.
\end{split}
\end{equation}

Equations (S30) and (S33) are Eq. (16) and (17) provided in the main text.

\end{document}